

\magnification 1200
\vsize=38pc
\hsize=34pc

\centerline{\bf GRASSMANN INTEGRAL TOPOLOGICAL INVARIANTS}

\vskip 3pc

\centerline{Ctirad Klim\v c\'\i k}

\centerline{\it Th. Div. of the Nuclear Centre, Charles University}
\centerline{\it Prague, Czechoslovakia}

\vskip 13pc

\leftline{\it Abstract:}

\vskip1pc

Partition functions of some two-dimensional statistical models can be
represented by means of Grassmann integrals over loops living
on two-dimensional torus. It is shown that those Grassmann integrals
are topological invariants,
 which depend only on the winding
numbers of the loops. The fact makes possible to evaluate the partition
functions of the models and the statistical
mean values of certain topological characteristics (indices) of
the configurations,
which behave as the (topological)  order parameters.

\vskip2pc

\vfill\eject

1. INTRODUCTION

\vskip 1pc

The integral topological invariants play important role in modern
mathematical physics. Among the most illustrative examples we may mention
the Pontryagin index, familiar from the gauge fields physics [8], or the
Euler characteristic of the manifold, entering the analysis of the string
perturbation theory [5]. Some physicists  tend to assign to
topological methods and ideas privileged role in the structure of
physical theories [10]. We do not wish to discuss  the importance
of topological methods in general, nevertheless, in this contribution, we
 present some
statistical models where topological considerations enter in a nontrivial
way. We shall study, in what follows, the partition functions of gases
of loops living on two-dimensional torus. In the cases studied, we can
express those partition functions by means of certain Grassmann integrals
over the loops.
The starting point of our treatment will be the Berezin's work [2], in which
the free fermionic representation of the partition function of the Ising
model in the plane was studied. Berezin has proved the lemma, that certain
Grassmann
integral over the loops has the same value for all nonselfintersecting loops
living in the plane. He remarked also, that the lemma does not hold, when
the plane
is replaced by the torus. On the other hand, quite recently it has arisen the
problem in the string statistics whether the nontrivial winding modes should
 be
summed over [3,7,9], when putting the strings in the
finite box with periodic boundary conditions. The present author has
constructed
the toy string model [6], in which the winding modes played a crucial role.
He has studied the gas of loops (or classical strings) living on the two
dimensional toroidal lattice and found the phase transition with the phases
differing by the parities of the winding numbers of the dominant
configurations.
In this contribution, we reproduce the results of [6] expressing the main
characteristics of the model by means of the
 Grassmann integrals introduced by
Berezin. The key ingredience of the method constitute in the generalization
of the Berezin's lemma to the case of the loops living on the torus.
We prove, in fact, that the value of the appropriately constructed Grassmann
integral over the loops is the function of the winding numbers of the
loops only, hence, it is the topological invariant. In particular, for all
loops in the plane the value of the integral is the same and we recover
the result of Berezin. The topological invariance of the constructed Grassmann
integral will enable us to evaluate the partition function and the
"topological"
order parameter" of the model. This order parameter measures the dominance
of the configurations with different parities as the function of the
temperature.
The equantity turns out to jump at the temperature of the phase transition.

The paper is organized as follows: In Sec.2 we introduce the Grassmann
integrals
over loops on the two-dimensional toroidal lattice and show their invariance
with respect to small deformations. Then we pick up a representative from each
class of loops with given winding numbers and evaluate the value of the
integral.
In Sec.3 we introduce the toy two-dimensional string model [6] and evaluate
its partition function. We also introduce and evaluate the topological order
 parameter,
mentioned above.

 We shall end up with short conclusions.

\vskip 1pc

2. GRASSMANN INTEGRALS OVER CLOSED LOOPS.

\vskip 1pc

Consider a square toroidal two-dimensional lattice with $N_1$ horizontal
and $N_2$ vertical sites. The distance between the neighbouring sites is
set to $1$ and the sites $(i_1,i_2)$ are parametrized by integers
$i_1\in(1,\dots,N_1)$ and $i_2\in(1,\dots,N_2)$. We assign to each site
four Grassmann variables $x_{i_1,i_2},x^*_{i_1,i_2},y_{i_1,i_2},y^*_{i_1,i_2}$
where each variable correspond to one half-link attached to the site in the
way described in Fig.1. We assign also to each pair of half-links attached
to the site the (ordered) product of the corresponding Grassmann variables
according Fig.2.

Draw now a closed nonselfintersecting loop $C$ on the lattice. We assign
it the Grassmann integral $I(C)$ as follows

$$I(C)=\int \prod_{links~ of~ C}{\rm d}\mu_{link} \prod_{sites~ of~ C}
A_r(site),~~~r \in (3,\dots,8)\eqno(2.1)$$

\noindent where the index $r$ denotes the way in which the loop runs over the
site
(see Fig.2) and the measure ${\rm d} \mu_{link}$ is given by

$$\rm d\mu_{link}\equiv {\rm d}x^*_{i_1+1,i_2}{\rm d}x_{i_1,i_2},\eqno(2.2a)$$

\noindent for the horizontal links and

$${\rm d}\mu_{link}\equiv {\rm d}y^*_{i_1,i_2+1}{\rm d}y_{i_1,i_2},
\eqno(2.2b)$$

\noindent for the vertical links.

We now show that the integral (2.1) is invariant for all loops with the
fixed horizontal and vertical winding numbers respectively.
 Consider first the case of
topologically trivial loops $C_{triv}$ e.g. those with both winding
numbers equal to zero. Each such loop can be regarded as the closed loop
in the "covering" plane of the torus (see Fig.3). The original torus is
obtained
by the identification of the points $(i_1 + k_1 N_1,i_2 + k_2 N_2)$ where
$k_1,k_2$ are integers. Since the loop is nonselfintersecting, the value of
 $I(C)$ is obviously unchanged
 in this
picture where to {\it each} point of the covering plane we associate its
own quadruple of the Grassmann variables. The closed loop in the plane
is the boundary of some domain. Now it is obvious that this loop can be
deformed by the succesion of small deformations to the boundary of a single
plaquette. Indeed, we can take away from the domain a plaquette after
plaquette until remains a single one. We do it in such a way that the removed
plaquette  participates on the nonempty connected piece of the boundary
of the domain (see Fig.4). It is a simple exercise to demonstrate that such
a plaquette always exists and that respecting this rule of removing the
plaquettes the boundary of each intermediate domain will consist of a single
loop. To prove the invariance of the integral $I(C_{triv})$ it is enough
to show that $I(C_{triv})$ does not change, removing a single plaquette from
the domain. This can be done by the simple inspection of all possibilities
(see Fig.5 and Lemma 1 of the Appendix for the details). The value of $I(C)$
is therefore
the same for all loops with both winding numbers equal to zero and is
given by $I(C_{plaquette})$ i.e.

$$\eqalign{&I(C_{plaquette})=\int {\rm d}y^*_{i_1,i_2 +1}
{\rm d}y_{i_1,i_2}{\rm d}x^*_{i_1+1,i_2+1}{\rm d}x_{i_1,i_2+1}
{\rm d}y^*_{i_1+1,i_2+1}{\rm d}y_{i_1+1,i_2}{\rm d}x^*_{i_1+1,i_2}
{\rm d}x_{i_1,i_2} \cr &\times y^*_{i_1,i_2+1}x_{i_1,i_2+1}x^*_{i_1+1,i_2+1}
y^*_{i_1+1,i_2+1}
x^*_{i_1+1,i_2}y_{i_1+1,i_2}x_{i_1,i_2}y_{i_1,i_2}=1=I(C_{triv})\cr}
\eqno(2.3)$$

\noindent We remind here the Berezin rules

$$\int x_{\alpha}{\rm d}x_{\alpha}=1,~~~~~~~~~\int {\rm d}x_{\alpha}=0,
\eqno(2.4)$$

$$\{x_{\alpha},x_{\beta}\}_+=\{x_{\alpha},{\rm d}x_{\beta}\}_+=
\{{\rm d}x_{\alpha},
{\rm d}x_{\beta}\}_+=0,\eqno(2.5)$$

\noindent where the (multi)index $\alpha$ denotes the variable (e.g.
$y^*_{i_1,i_2}$).

The loop $C_{k,l}$ with (at least one nonzero) winding numbers $[k,l]$
also can be regarded as the contour in the covering plane, starting at
$(0,0)$ and ending at $(kN_1,lN_2)$. We can compute easily
$I(C_{k,l})$ as follows. First we perform explicitly the integration
over the star-variables in (2.1) which gives

$$I(C)=\int \prod_{sites ~of~C} B_r(site),\eqno(2.6)$$

\noindent where $B_r$ are associated with the pairs of half-links entering
a site
according the Fig.2 and are given by

$$~B_3={\rm d}x_{i_1-1,i_2}x_{i_1,i_2},~~~B_4={\rm d}y_{i_1,i_2-1}
y_{i_1,i_2},~~~B_5=
{\rm d}x_{i_1-1,i_2}y_{i_1,i_2},$$

$$B_6={\rm d}y_{i_1,i_2-1}x_{i_1,i_2},~~~B_7={\rm d}x_{i_1-1,i_2}{\rm d}
y_{i_1,i_2-1},~~~B_8=x_{i_1,i_2}y_{i_1,i_2},\eqno(2.7)$$

\noindent Then we realize that each nonselfintersecting contour $C_{k,l}$
can be
completed to a closed nonselfintersecting loop $C^c_{k,l}$ in the covering
plane in such a way, that the part, which completes the original contour,
is the
union of three straight lines (see Fig.8 and, for the proof, Lemma 5 of the
Appendix). For concreteness, we shall work with the case $l\neq 0$, in which
two of three completing lines are horizontal and they join the original
contour
at the vertex of the type $A_6$ (see Fig.8). Now denote $C^+_{k,l}$ the piece
of
$C^c_{k,l}$ which completes $C_{k,l}$ in the covering plane. We can write (see
Fig.8)

$$\eqalign{&I(C^c_{k,l})=\int \big[\prod_{sites~of~C^c_{k,l}~besides~(0,0),
(kN_1,lN_2)}B_r(site)\big]{\rm d}x_{-1.0}x_{0,0}{\rm d}x_{kN_1-1,lN_2}
{\rm d}y_{kN_1,lN_2-1}\cr &= \int\big[\prod_{sites~of C^+_{k,l}}B_r \big]
\big[\prod_{sites~ of~ C_{k,l}~ besides ~(0,0),(kN_1,lN_2)}B_r \big]
{\rm d}x_{-1,0}{\rm d}x_{kN_1-1,lN_2}{\rm d}y_{kN_1,lN_2-1}x_{0,0}\cr &=
I(C_{k,l})\times \int \big[\prod_{sites~of~C^+_{k,l}}B_r(site)\big]
{\rm d}x_{-1,0}
{\rm d}x_{kN_1-1,lN_2}=1\cr} \eqno(2.8)$$

\noindent Hence

$$I^{-1}(C_{k,l})=\int\big[\prod_{sites~of~C^+_{k,l}}B_r(site)\big]
{\rm d}x_{-1,0}
{\rm d}x_{kN_1-1,lN_2}=-1.\eqno(2.9)$$

\noindent All remaining cases ($l=0$ or joining the original contour at the
vertex of different type)
can be treated in full analogy with this one and
we get
 for all $[k,l]$, except $[0,0]$,

$$I(C_{k,l})=-1.\eqno(2.10)$$

\noindent We conclude with the formula valid for all $[k,l]$ i.e.

$$I(C_{k,l})=(-1)^{k+l+kl},\eqno(2.11)$$

\noindent (2.11) holds due to the fact that nonselfintersecting loops on the
toroidal
lattice with both winding numbers even are necessarily topologically trivial.
The proof of this statement we present in the Lemma 2 of the Appendix.

\par\vfill\supereject

3. THE TOPOLOGICAL ORDER PARAMETER.

\vskip1pc

Consider the following Grassmann integral, defined in the Grassmann algebra
associated with the toroidal lattice in the way described in the preceding
section

$$S_{N_1,N_2}(\rho'_r)=\int \big(\prod_{sites}\sum^8_{r=1}\rho'_r A_r(site)
\big)
\times \exp{\sum_{i_1,i_2}[x_{i_1,i_2}x^*_{i_1+1,i_2}+y_{i_1,i_2}
y^*_{i_1,i_2+1}]}
\prod_{links}{\rm d}\mu_{link}\eqno(3.1)$$

\noindent where we set

$$A_1=A_7A_8,~~~~~~A_2=1    \eqno(3.2)$$

\noindent The numbers $\rho'_r$ are taken from the interval $<0,1>$. Expanding
the product in (3.1) we get the sum of the terms of the type

$$\prod_{i_1,i_2}A_{r(i_1,i_2)}\rho'_{r(i_1,i_2)} \eqno(3.3)$$

\noindent Each such term can be graphically represented as the lattice with
the vertices
$A_r$ marked according Fig.2. Clearly not all such terms will contribute to
$S_{N_1,N_2}(\rho'_r)$. Indeed, if there is a link with just one  marked half,
the Grassmann variable corresponding to the other half of the link is absent
in the integrand and, due to Berezin rules (2.4), the term possessing such a
link
gives zero contribution. Now it is easy to see, that terms, the graphical
representation of which do not have the halfoccupied links, do contribute
to $S_{N_1,N_2}(\rho'_r)$, because if there is a link with no  marked half,
the corresponding pair of the Grassmann variables sitting at the exponent
in (3.1) makes the contribution nonzero. As the example of the graphical
representation of a contributing term, we can take Fig.4 in which the vertex
$A_1$ is drawn as the union of the vertices $A_7$ and $A_8$. Such a rule of
drawing $A_1$ enables us to consider the contributing terms as the weighted
configurations of nonselfintersecting  and mutually nonintersecting loops
living on two dimensional toroidal lattice. Moreover,we can write for
$S_{N_1,N_2}(\rho'_r)$

$$S_{N_1,N_2}(\rho'_r)=\sum_{configurations~ of~ loops}\big(\prod_{r=1}^8
\rho_r^{'a_r}\big)
\prod_j I(C_j) \eqno(3.4)$$

\noindent where $a_r$ are numbers of $r$-vertices in the configuration,
$I(C)$ is the
integral (2.1) and pieces of loops meeting each other at one vertex should
avoid each other in the way depicted in Fig.4. If $I(C)$ were always $+1$,
$S_{N_1,N_2}(\rho'_r)$ would give us precisely the partition function of the
eight-vertex model  with the Boltzmann weights
$\rho'_r$(generally in an external field) [1]. As we see from (2.11), the
noncontractible loops spoil the interpretation
of $S_{N_1,N_2}(\rho'_r)$ as the statistical partition function of some model,
since there are configurations  entering the sum with  negative weights.
 We can save the day,
however, as follows. Take odd the both dimensions $N_1$ and $N_2$ of the
lattice
and set

$$\eqalign{&\rho'_1=\lambda_1\lambda_2\rho_1,~~~\rho'_2=\rho_2,~~~\rho'_3=
\lambda_1\rho_3,
{}~~~,\rho'_4=\lambda_2\rho_4,\cr &\rho'_5=\lambda_2\rho_5,~~~\rho'_6=
\lambda_1\rho_6,
{}~~~\rho'_7=\lambda_1\lambda_2\rho_7,~~~\rho'_8=\rho_8.\cr}\eqno(3.5)$$

\noindent Then

$$\big(\prod_{r=1}^8\rho_r^{'a_r}\big)\prod_j I(C_j)=
\big(\prod_{r=1}^8\rho_r^{a_r}\big)
\prod_j\big[I(C_j) \lambda_1^{h_j}\lambda_2^{v_j}\big], \eqno(3.6)$$

\noindent where $h_j$ and $v_j$ are the numbers of  the horizontal and the
vertical links of the $j$-th loop respectively. If $N_1$ and $N_2$ are odd
then

$$(-1)^{h_j}=(-1)^{k_j},\eqno(3.7a)$$

$$(-1)^{v_j}=(-1)^{l_j},\eqno(3.7b)$$

\noindent where $[k_j,l_j]$ are the winding numbers
of the loops, respectively. Then we realize (for the proof see Lemma 4 of the
Appendix),
that  for the configuration of nonselfintersecting and mutually
nonintersecting
loops $C_j$ on the lattice, it holds

$$\prod_j I(C_j)= (-1)^{\sum k_j +\sum l_j +(\sum k_j)(\sum l_j)}\eqno(3.8)$$

\noindent Define

$$Z_{N_1,N_2}(\lambda_1,\lambda_2)\equiv {1\over 2}[S_{N_1,N_2}
(-\lambda_1,-\lambda_2)+
S_{N_1,N_2}(-\lambda_1,\lambda_2) +S_{N_1,N_2}(\lambda_1,-\lambda_2)-
S_{N_1,N_2}(\lambda_1,\lambda_2)]\eqno(3.9)$$

\noindent Combining (3.6),(3.7) and (3.8), it follows

$$Z_{N_1,N_2}(\lambda_1=1,\lambda_2=1)=\sum_{configurations~of~loops}
\big(\prod_{r=1}^8\rho_r^{a_r}\big),\eqno(3.10)$$

\noindent or, in other words, the combination (3.9) of $S_{N_1,N_2}
(\pm \lambda_1,\pm \lambda_2)$
gives the partition function of the eight-vertex model.
Without loss of generality, setting $\rho_2=1$ we may rewrite (3.1) as follows

$$\eqalign{S_{N_1,N_2}(\rho'_r)&=\int \exp{\big[\sum_{sites}
\big(\sum_{r=3}^8 \rho'_r
A_r(site) + (\rho'_1 +\rho'_3 \rho'_4 -\rho'_5\rho'_6 -\rho'_7\rho'_8)
A_7(site) A_8(site)\big)\big]}\cr &\times\exp{\big[\sum_{sites}
(x_{i_1,i_2}x^*_{i_1+1,i_2}+
y_{i_1,i_2}y^*_{i_1,i_2+1})\big]}\prod_{links}{\rm d}\mu_{link}\cr}
\eqno(3.11)$$

\noindent Note, that if

$$\rho'_1=\rho'_5\rho'_6 +\rho'_7\rho'_8-\rho'_3\rho'_4,\eqno(3.12)$$

\noindent the expression in the exponent is the quadratic form in the
Grassmann algebra
with the cyclic matrix  and , therefore, $S_{N_1,N_2}(\rho'_r)$ can be
computed
easily in this case. The constraint (3.12) is well-known
 and gives the so-called free fermionic sector of the eight-vertex model
[1]. If (3.12) does not hold, we have the quartic term in the exponent,
hence,  we may
call the formulas (3.11),(3.5) and (3.9) interacting fermionic representation
of the
eight-vertex model.

In what follows we shall study the particular case\footnote{*}{ This
"six-vertex"
model differs from the usual one, in which $\rho_7=\rho_8=0$}
in which

$$\rho_3=\rho_4=0,~~~\rho_2=\rho_5=\rho_6=\rho_7=\rho_8=1,~~~\rho_1=2,
\eqno(3.13)$$

$$\lambda_1=\lambda_2={\rm e}^{-\beta}.\eqno(3.14)$$

\noindent (3.12)
is
obviously satisfied, therefore we can easily compute $S_{N_1,N_2}(\rho'_r)$.
Before doing that, however, let us look more closely on the set of
 configurations
when conditions (3.13) apply. The vertices $A_3$ and $A_4$ are absent, thus
the loops
locally have a "zig-zag" shape i.e. at every vertex of the lattice the loop
has to change
its direction. Moreover, the vertex $A_1$ enters with the weight $\rho_1=2$.
Without endangering our results, in this particular case, we may  abandon our
(conventional) way of drawing the vertex $A_1$, as described in Fig.4. The
only aim of this convention constituted in representing the contributing
configurations
to $S_{N_1,N_2}(\rho'_r)$ as the configurations of nonintersecting loops.
If the vertex $A_1$
itself enters with the weight $\rho_1=2$, we can view it
 as corresponding to two ways of connecting the half-links
of the $A_1$ vertex (see Fig.6). The factor ${\rm e}^{-\beta}$ causes
that the weight
$w_c$, which the loop enters the sum with, is given by

$$w_c ={\rm e}^{-L_c \beta},\eqno(3.15)$$

\noindent where $L_c$ is the lenght of the loop. Indeed, each link is
accompanied by one of the factors $\lambda_1$ or $\lambda_2$, as it follows
from (3.5). Summarizing, we may look at the particular case (3.13-14) as at
the lattice
regularization of the model of free classical strings (loops) with energies
proportional to their lenghts and with the configurations $6_a$ and $6_b$
considered
as different ( corresponding to splitting and joining of strings).

The free fermionic representation (3.9) and (3.11) of the partition function
of such string model and the value (2.11) of the Grassmann integral
topological
invariant make possible to evaluate the
statistical mean values of some topological characteristics of the strings
such as the functions of the winding numbers are. In our case, we actually
have a constraint
in the space of loops, since the lattice regularization respects the zig-zag
rule
and the dimensions $N_1$ and $N_2$ of the lattice are both odd. In fact, the
zig-zag rule means that

$$h=v,\eqno(3.16)$$

\noindent where $h(v)$ is the horizontal (vertical) lenght of the string,
respectively. Combining this fact with (3.7), we see

$$(-1)^k =(-1)^l, \eqno(3.17)$$

\noindent where $k(l)$ is the horizontal (vertical) winding number of the
string,
respectively. Therefore, the parities of both winding numbers of the string
are  the same.

In what follows, we shall calculate the quantity

$$<(-1)^K>_{\beta,V\rightarrow\infty},\eqno(3.18)$$

where $V\equiv N_1N_2$ is the volume of the system and $K$ is the total
horizontal winding number of the configuration of strings.\footnote{**}{The
strings are nonoriented, nevertheless $(-1)^K$ can be defined unambiguously
assigning to each string whatever orientation.}
It is given by

$$<(-1)^K>={\sum_{configurations}(-1)^{\sum k_j} {\rm e}^{-\beta\sum L_j}
\over
\sum_{configurations}{\rm e}^{-\beta\sum L_j}}\eqno(3.19)$$

\noindent where $k_j(L_j)$ is the horizontal winding number (total lenght) of
the
$j$-th string of the configuration.  In the case of our "zig-zag" string
model (3.8)
and (3.17) imply

$$\prod_j I(C_j)=(-1)^{\sum k_j}, \eqno(3.20)$$

\noindent hence, following (3.4), we may write

$$<(-1)^K>={S_{N_1,N_2}(\lambda_1={\rm e}^{-\beta},\lambda_2 ={\rm e}^{
-\beta})
\over Z_{N_1,N_2}(\lambda_1 ={\rm e}^{-\beta},\lambda_2 ={\rm e}^{-\beta})},
\eqno(3.21)$$

\noindent where $Z_{N_1,N_2}(\lambda_1,\lambda_2)$ is the partition function
of the model and it is given by (3.9). In our "zig-zag" case

$$S_{N_1,N_2}(-\lambda_1,\lambda_2)=S_{N_1,N_2}(\lambda_1,-\lambda_2),
\eqno(3.22a)$$

$$S_{N_1,N_2}(-\lambda_1,-\lambda_2)=S_{N_1,N_2}(\lambda_1,\lambda_2),
\eqno(3.22b)$$

\noindent as it follows from (3.4),(3.6) and (3.16). Using (3.9) then we have

$$Z_{N_1,N_2}(\lambda_1,\lambda_2)=S_{N_1,N_2}(-\lambda_1,\lambda_2)
\eqno(3.23)$$

\noindent and

$$<(-1)^K>={S_{N_1,N_2}(\lambda_1={\rm e}^{-\beta},\lambda_2={\rm e}^{-\beta})
\over
S_{N_1,N_2}(\lambda_1=-{\rm e}^{-\beta},\lambda_2={\rm e}^{-\beta})}.
\eqno(3.24)$$

Actual computation of $S_{N_1,N_2}$ is easy, since the model fulfils the free
fermionic constraint (3.11). Using the discrete Fourier transformation we can
easily diagonalize the matrix of the quadratic form sitting at the exponent of
(3.10) and obtain

$$S_{N_1,N_2}(-{\rm e}^{-\beta},{\rm e}^{-\beta}) = (1+2{\rm e}^{-2\beta})
\prod_{(p,q)\neq(0,0)}(4{\rm e}^{-4\beta}+1+4{\rm e}^{-2\beta}
\cos{{2\pi\over N_1}p}
\cos{{2\pi\over N_2}q})^{{1\over 2}} \eqno(3.25a)$$

\noindent and

$$S_{N_1,N_2}({\rm e}^{-\beta},{\rm e}^{-\beta})=(1-2{\rm e}^{-2\beta})
\prod_{(p,q)\neq(0,0)}(4{\rm e}^{-4\beta}+1-4{\rm e}^{-2\beta}
\cos{{2\pi\over N_1}p}
\cos{{2\pi\over N_2}q})^{{1\over 2}}, \eqno(3.25b)$$

\noindent where the square roots in (3.25) should be taken positive. Note
that at
certain critical inverse temperature

$$\beta_c = {\rm ln}\sqrt{2} \eqno(3.26)$$

\noindent $S_{N_1,N_2}({\rm e}^{-\beta},{\rm e}^{-\beta})$ changes its sign!
Moreover, the free energy per site of the model

$$F(\beta)=1-{1\over 2\pi\beta}\int_0^{\pi}{\rm d}x {\rm Arch}[{{\rm cosh}
(2\beta-{\rm ln}2)\over \vert \cos{x}\vert}]\eqno(3.27)$$

\noindent is nonanalytic at $\beta_c$ and we have the second-order phase
transition in the
system.

The computation of $<(-1)^K>$ is not difficult. We use the formulas [4,6]

$$2^{n-1}\prod_{r=0}^{r=n-1}\cos{(\theta+{2\pi\over n}r)}=\cos{n\theta},
\eqno(3.28a)$$

$$2^{n-1}\prod_{r=0}^{r=n-1}\{\cosh{\phi} - \cos{(\theta+{2\pi\over n}r)}\}=
\cosh{n\phi}-\cos{n\theta}\eqno(3.28b)$$

\noindent and write

$$\eqalign{&S_{N_1,N_2}(-{\rm e}^{-\beta},{\rm e}^{-\beta})=
{\rm e}^{-\beta N_1 N_2}2^{N_2 +{N_1\over 2}}
\cosh{N_1(\beta-{\rm ln}\sqrt{2})}\cr
&\prod_{{2\pi\over N_2}q\in (0,{\pi\over 2})}
{\rm cosh}^2\big[{N_1\over 2}{\rm Arch}\big({\cosh{(2\beta-{\rm ln}2)}\over
\vert \cos{{2\pi\over N_2}q}\vert}\big)\big]
\prod_{{2\pi\over N_2}q\in ({\pi\over 2},\pi)}
{\rm sinh}^2\big[{N_1\over 2}{\rm Arch}\big({\cosh{(2\beta-{\rm ln}2)}\over
\vert \cos{{2\pi\over N_2}q}\vert}\big)\big]\cr}\eqno(3.29a)$$

$$\eqalign{&S_{N_1,N_2}({\rm e}^{-\beta},{\rm e}^{-\beta})=
{\rm e}^{-\beta N_1 N_2}2^{N_2 +{N_1\over 2}}
\sinh{N_1(\beta-{\rm ln}\sqrt{2})}\cr
&\prod_{{2\pi\over N_2}q\in (0,{\pi\over 2})}
{\rm sinh}^2\big[{N_1\over 2}{\rm Arch}\big({\cosh{(2\beta-{\rm ln}2)}\over
\vert \cos{{2\pi\over N_2}q}\vert}\big)\big]
\prod_{{2\pi\over N_2}q\in ({\pi\over 2},\pi)}
{\rm cosh}^2\big[{N_1\over 2}{\rm Arch}\big({\cosh{(2\beta-{\rm ln}2)}\over
\vert \cos{{2\pi\over N_2}q}\vert}\big)\big]\cr}\eqno(3.29b)$$

\noindent Dividing (3.29a) by (3.29b) we get from (3.24)

$$<(-1)^K>_{\beta,N_1\to\infty}={\rm sign}(\beta-{\rm ln}\sqrt{2}),
\eqno(3.30a)$$

\noindent hence

$$<(-1)^K>_{\beta,V\to\infty}={\rm sign}(\beta-{\rm ln}\sqrt{2}).
\eqno(3.30b)$$

\noindent  We observe that $<(-1)^K>$ can be interpreted as the topological
order parameter which says that at low
temperatures the "even" configurations dominate while at high temperatures
the "odd" (and necessarily topologically nontrivial) configurations are
dominant.
Note that from the technical point of view, we could obtain the result
particularly due to the
invariant character of the integral (2.1) and due to formula (2.11).

\vskip1pc

4. CONCLUSIONS.

\vskip1pc

In this contribution, we have constructed and evaluated the Grassmann
integral topological invariants and found the applications of these
results in the
field of statistical physics. In particular, we were able to introduce and
to evaluate the topological order parameter in the toy string model
constructed
by the author previously in [6]. We believe that our results can be
generalized in two
directions. First one, more mathematical, would constitute in further study
of the Grassmann integrals, either with more complicated integrands or
over the
loops living on more complicated manifolds. The second direction would
constitute in studying the existence and the behaviour of the topological
order parameters in various physical systems. We intend to pursue these
problems with the hope of obtaining new interesting results.

\vskip1pc

\par\vfill\supereject

APPENDIX.

\vskip1pc

{\bf Lemma 1}:{\it  Small deformations of loops in the plane, induced
by removing a single plaquette from the domain bounded by the loop, do
not change the value of the integral (2.1), provided the removed plaquette
participates on the nonempty connected piece of the boundary of the domain.}

\vskip1pc

{\it Proof}: In Fig.5a-e we list all possibilities of removing the plaquette
in the way described in the formulation of the lemma. The plaquette to be
removed is depicted. The term "rotations" in Fig.5 means three
other possible orientations of the drawing rotated by the multiple of
 ${\pi\over 2}$
with respect to the depicted one. The new deformed loop follows the dashed
line. Both original and deformed loops have (large) common part alluded by
the half-links. In Fig.5b,c,d there are two half-links attached to  each
"connecting"
vertex indicating two possible ways of the continuation of the loop from
the vertex. The proof of the lemma is performed by the simple inspection
of all possibilities. For concreteness, we present the proof for the case
of Fig.5b in the original (i.e. nonrotated) orientation and with both
continuing
half-links pointing to the horizontal directions. We use formulae (2.6),(2.7)
and write

$$\eqalign{&I(C_{original})=\cr &=
\int\dots({\rm d}x_{i_1-1,i_2}{\rm d}y_{i_1,i_2-1})
(x_{i_1,i_2-1}y_{i_1,i_2-1})({\rm d}x_{i_1,i_2-1}y_{i_1+1,i_2-1})
({\rm d}y_{i_1+1,i_2-1}x_{i_1+1,i_2})\dots\cr&=\int\dots{\rm d}x_{i_1-1,i_2}
x_{i_1+1,i_2}\dots=\int\dots({\rm d}x_{i_1-1,i_2}x_{i_1,i_2})
({\rm d}x_{i_1,i_2}x_{i_1+1,i_2})\dots=I(C_{deformed})\cr}\eqno(A.1)$$

\noindent where the dots stand instead of terms common to both loops.
Analogously we can verify all remaining cases, thus proving the lemma.

\vskip1pc

{\bf Lemma 2}: {\it Let $C$ be the nonselfintersecting noncontractible
 loop on the toroidal lattice
of the type $[am,an],a\in Z$. Then $\vert a \vert =1$. In particular, the
nonselfintersecting loops on the toroidal lattice with both winding
numbers
even are necessarily topologically trivial.}

\vskip1pc

{\it Proof}: Consider the nonselfintersecting noncontractible loop $C$ on
the toroidal lattice,
with the winding numbers $[k,l]$. To each point $P_0
\equiv (p_1N_1,p_2N_2);p_1,p_2\in
Z$ we associate the (infinitely long) "covering" contour $C_{cov}(P)$, which
connects the point $P_b\equiv (p_1N_1+bkN_1,p_2N_2 +blN_2)$ with the point
$P_{b+1}\equiv (p_1N_1+(b+1)kN_1,p_2N_2 +(b+1)lN_2)$ for all integer $b$ and,
 for a given
$b$, $C_{cov}(P)$ is given by the canonical mapping of the loop $C$ to the
covering plane. The contour $C_{cov}(P)$ divides the covering plane in two
pieces
$L_+(C_{cov}(P))$ and $L_-(C_{cov}(P))$ defined as follows (see Fig.7)

$$\displaystyle\lim_{t\to\pm\infty}{(-t l N_2,t k N_1)}\in L_{\pm}
(C_{cov}(P)),~~~t\in Z. \eqno(A.2)$$

The piece of the contour $C_{cov}(P)$ between the points $P_b$ and $P_{b+1}$
is
finite, hence the contour $C_{cov}(P)$ is necessarily contained in the strip,
which
is finitely thick in the transverse direction given by the "normal" vector
$(-lN_2,kN_1)$. The definition (A.2), therefore, is selfconsistent.

Let $P\equiv (p_1N_1,p_2N_2),Q\equiv(q_1N_1,q_2N_2))$ be two points in the
covering plane. We show that

$$Q\in L_{\pm}(C_{cov}(P))\Longleftrightarrow P\in L_{\mp}(C_{cov}(Q)).
\eqno(A.3)$$

Indeed, if $Q\in L_{\pm}(C_{cov}(P))$ then $C_{cov}(Q)
\subset L_{\pm}(C_{cov}(P))$, since
the loop $C$ is nonselfintersecting. Then, obviously, it follows
$C_{cov}(P)\subset L_{\mp}(C_{cov}(Q))$, hence $P\in L_{\mp}(C_{cov}(Q))$.
In complete analogy the inverse implication is valid.

Define

$$P>_{C}Q \Longleftrightarrow Q\in L_{-}(C_{cov}(P)),\eqno(A.4a)$$

$$P=_{C}Q\Longleftrightarrow Q\in C_{cov}(P),\eqno(A.4b)$$

$$P<_{C}Q\Longleftrightarrow Q\in L_{+}(C_{cov}(P)).\eqno(A.4c)$$

\noindent Obviously $P>_{C}Q$ and $Q>_C R$ implies $P>_C R$, since
$R\in L_{-}(C_{cov}(Q))$ and \hfill

\noindent $L_{-}(C_{cov}(Q))\subset L_{-}(C_{cov}(P))$.

Now consider the nonselfintersecting noncontractible loop $C$ with
the winding numbers
$[k=am,l=an],a\in Z$, the points $P_0=(0,0)$,$P_1=
(\vert a\vert mN_1,\vert a\vert nN_2)$,
$Q_1=(mN_1,nN_2)$,
 $Q_2=(2mN_1,2nN_2),\dots,Q_{\vert a\vert -1}=((\vert a\vert -1)mN_1,
 (\vert a\vert -1)nN_2)$
and the covering contour $C_{cov}(P_0)$. Suppose $P_0 >_C Q_1$.
 Due to invariance
with respect to the shifts of the covering plane induced by the vectors
$(rN_1,sN_2);r,s\in Z$, we have $Q_1>_C Q_2>_C \dots >_{C} Q_{a-1}>_C P_1$.
Hence $Q_1>_C P_1$. On the other hand, $C_{cov}(P_0)\equiv C_{cov}(P_1)$,
which,
together with (A.3), implies $P_0<_C Q_1$ and we ended up with the
contradiction.
Analogously, the assumption $P_0<_C Q_1$ leads to the contradiction
$P_0 >_C Q_1$.
It remains the possibility $P_0=_C Q_1$. In this case $Q_1\in C_{cov}(P_0)$,
hence, since $C$ is nonselfintersecting, $Q_1\equiv P_1$ and the loop has
the winding
numbers $[am,an]$ where $\vert a\vert =1$.

\vskip1pc

{\bf Lemma 3}: {\it Let $C$ be a noncontractible loop of the type $[k,l]$.
Let $P\in C,
P\equiv (0,0)$ be the point in the covering plane of the torus; $t, p,q,x,y$
are integers. Then (for the notation see the proof of Lemma 2)

$$\displaystyle\lim_{t\to+\infty}{(t p N_1 +x,t q N_2 +y)}\in L_{\pm}
(C_{cov}(P))
\Longleftrightarrow \displaystyle\lim_{t\to -\infty}{(t p N_1, t q N_2)}
\in L_{\mp}(C_{cov}(P)), \eqno (A.5)$$

\noindent unless $[p,q]=[ck,cl];c\in Z$. In other words, following the line
$(t p N_1 +x,t q N_2 +y)$, where $t$ varies, we have to connect two
pieces $L_{+}(C_{cov}(P))$ and $L_{-}(C_{cov}(P))$ of the covering plane,
unless
$[p,q]=[ck,cl]$.}

\vskip1pc

{\it Proof}: The contour $C_{cov}(P)$ runs through the points
$(bkN_1,blN_2);b\in Z$.
Between the points $(bkN_1,blN_2)$ and $((b+1)kN_1,(b+1)lN_2)$ the
lenght of the
contour is finite, since the original loop $C$ , living on the torus,
 has finite lenght.
That means that the strip $S$ exists in the covering plane, containing
$C_{cov}(P)$
and dividing the covering plane $L$ in three pieces i.e. $S,S_+$ and
$S_-$, such
that $S_{\pm}\subset L_{\pm}(C_{cov}(P))$. Since the boundary lines between
$S_+$ and $S$ and between $S$ and $S_-$  have the tangent vector $[kN_1,
lN_2]$,
the line with the tangent vector $[pN_1,qN_2]$ has to connect $S_+$ with $S_-$
(and, therefore, $L_{+}(C_{cov}(P))$ with $L_-(C_{cov}(P))$) unless
 $[p,q]=[ck,cl]$.

\vskip1pc

{\bf Lemma 4}: {\it For the configuration of nonselfintersecting and mutually
nonintersecting
loops $C_j$ on the toroidal lattice with the winding numbers $[k_j,l_j]$,
 it holds}

$$\prod_j I(C_j) = (-1)^{\sum k_j + \sum l_j +(\sum k_j)(\sum l_j)}.
\eqno(A.6)$$

\vskip1pc

{\it Proof}: If at most one loop of the configuration is noncontractible,
the proposition obviously holds. Now let $C_1$ and $C_2$ be two
noncontractible
mutually nonintersecting loops at the toroidal lattice, of the types
 $[k_1,l_1]$
and $[k_2,l_2]$ respectively. Take two points $P_1\in C_1$ and $P_2\in C_2$
and
lift them to the fundamental domain of the covering plane. Consider then two
 contours $C_{1,cov}(P_1)$ and $C_{2,cov}(P_2)$. Without loss of generality,
we set $P_1=(0,0),P_2=(x,y)$. The contour $C_{2,cov}(P_2)$ runs through the
points $(bk_2 N_1 +x, bl_2N_2 +y);b\in Z$. Following Lemma 3, unless
$[k_2,l_2]=[ck_1,cl_1],c\in Z$, it has to connect $L_+(C_{1,cov}(P_1))$ with
$L_-(C_{1,cov}(P_1))$ and, consequently, to intersect $C_{1,cov}(P_1)$.
Since the loops are nonintersecting, this means $[k_2,l_2]=[ck_1,cl_1]$.
{}From Lemma 2 it follows, however, that $\vert c \vert=1$. In conclusion,
the loops $C_1$ and $C_2$ have the same winding numbers (up to the sign).
Analogously it can be shown that $n$ mutually nonintersecting noncontractible
loops on the toroidal lattice must have the same winding numbers. The formula
(A.6)
then trivially follows.

\vskip1pc

{\bf Lemma 5}: {\it Each noncontractible nonselfintersecting loop $C$ on the
toroidal lattice can be completed  to the closed nonselfintersecting
loop in the covering plane, in such a way that the part, which completes
the original contour, is the union of three straight lines (see Fig.8).}

\vskip1pc

{\it Proof}: Consider the loop $C$ of the type $[k,l];l\neq 0$. Following
Lemma 3,
each (infinite) horizontal line has to intersect $C_{cov}(P)$. Pick up two
horizontal lines, $H_0$ and $H_1$, with the mutual vertical distance equal
to $N_2$.
Travelling from the left these lines intersect $C_{cov}(P)$ for the first
time at the
points $B_0$ and $B_1$, respectively. Now there exist two points $A_0$ and
$A_1$,
belonging to the first and the second line, respectively, such that they both
lie to the left from the points $B_0$ and $B_1$, respectively, their
horizontal coordinate is the same and the line $A_0-A_1$ does not intersect
$C_{cov}(P)$.
 The loop $C^c$ connecting the points
$A_0-A_1-B_1-B_0-A_0$ is the closed nonselfintersecting loop in the covering
plane,
with the required properties. If $l=0$, we take $V_0$ and $V_1$ to be the
vertical lines with the mutual horizontal distance equal $N_1$ and construct
the
loop $C^c$ in the full analogy with the previous case.

\par\vfill\supereject

FIGURE CAPTIONS.

\vskip2pc

\noindent Fig.1: The association of the Grassmann variables to the halflinks
of the lattice.

\vskip1pc

\noindent Fig.2: The association of the (ordered) products of the Grassmann
variables to each even

subset of the halflinks attached to a single site of the lattice.

\vskip1pc

\noindent Fig.3: The covering plane of the torus. The rectangle $(0,0)-
(N_1,0)-(N_1,N_2)-(0,N_2)$ is

the fundamental domain. Other rectangles
are its copies.

\vskip 1pc

\noindent Fig.4: The illustration of the allowed way of the deformations
of loops by removing the

plaquettes. The plaquette marked by "n"
participates on the nonempty {\it disconnected} piece

of the boundary of the
domain, hence, it must not to be removed. The plaquette "y",

instead, can be
removed yielding the deformed loop 4b. Note the way of drawing the

vertex of the type
$A_1$.

\vskip1pc

\noindent Fig.5: The ways of deforming the loops by removing a single
plaquette. The plaquette to

be removed is depicted. The term "rotations"
means three other possible orientations of

the drawing, rotated by the multiple
of ${\pi\over 2}$ with respect to the depicted one. The

new deformed loop
follows the dashed line. Both original and deformed loops

have (in general
large) common part, alluded by the halflinks. In Fig.b,c,d there are

two
halflinks attached to each "connecting" vertex, indicating two possible ways
of the

continuation
of the loop from the vertex.

\vskip1pc

\noindent Fig.6: Two ways of connecting the halflinks of the $A_1$ vertex,
corresponding
to splitting and

joining of strings.

\vskip1pc

\noindent Fig.7:
The illustration of the construction of the contour $C_{cov}(P)$. The contour
divides the

covering plane in two pieces, marked by $L_+$ and $L_-$. The
"normal" line to $C_{cov}(P)$ is also

depicted. Both winding numbers are chosen to be positive.

\vskip1pc

\noindent Fig.8: Completing of the contour in the covering plane by the union
of three straight lines.

\par\vfill\supereject

REFERENCES.

\vskip2pc

\noindent [1] Baxter, R.: Exactly Solved Models in Statistical Mechanics.
New York: Academic Press 1982

\vskip1pc

\noindent [2] Berezin,F.: Plane Ising model. Uspekhi mat.nauk {\bf 24},3-22
(1969)

\vskip1pc

\noindent [3] Brandenberger, R., Vafa, C.: Superstrings in the
early universe. Nucl.Phys. {\bf B316}, 391-410 (1989)

\vskip1pc

\noindent [4] Gradshteyn, I.S., Ryzhik, I.M.: Tables of Integrals, Summs,
Series and Products. Moscow: State Publishing House for Phys. and Math.
Literature,

1963

\vskip1pc

\noindent [5] Green, M., Schwarz, J., Witten,E.: Superstring Theory.
Cambridge: Cambridge Press 1987

\vskip1pc

\noindent [6] Klim\v c\'{\i}k, C.: Thermal strings on the lattice and the
winding

number transition. Phys.Lett. {\bf B263}, 469-475 (1991)

\vskip1pc

\noindent [7] Mitchell, D., Turok, N.: Statistical mechanics of cosmic
strings. Phys.Rev.Lett. {\bf 58}, 1577-1580 (1987),

Statistical properties of cosmic strings. Nucl.Phys. {\bf B294}, 1138-1163
(1987)

\vskip1pc

\noindent [8] Rajaraman, R.: Solitons and Instantons. Amsterdam:
North-Holland, 1982

\vskip1pc

\noindent [9] Salomonson, P., Skagerstam, B.S.: On superdense superstring
gases. Nucl.Phys. {\bf B268}, 349-361 (1986)

\vskip1pc

\noindent [10] Witten,E.: talk at the Summer School, Trieste 1989

\vskip1pc

\par\vfill\supereject

\end

\end

\end